\documentclass[twocolumn,pre]{revtex4}
\usepackage{epsfig}
\usepackage{amssymb,amsfonts,amsmath}
\usepackage[pdftex]{pstricks}
\include{epsf}

\newcommand{\bra}[1]{\langle{#1}|}
\newcommand{\ket}[1]{|{#1}\rangle{}}
\newcommand{\braket}[2]{\langle{}{#1}|{#2}\rangle{}}
\newcommand{\<}{\langle}
\renewcommand{\>}{\rangle}

\newcommand{\beq}{\begin{equation}}
\newcommand{\eeq}{\end{equation}}
\newcommand{\bea}{\begin{eqnarray}}
\newcommand{\eea}{\end{eqnarray}}

\begin{document}
\title{Limits of sensing temporal concentration changes by single cells}

\author{Thierry Mora}

\address{Lewis-Sigler Institute for Integrative Genomics,
Princeton University, Princeton, New Jersey, USA}

\author{Ned S. Wingreen}

\address{Department of Molecular Biology,
Princeton University, Princeton, New Jersey, USA}


\begin{abstract}
Berg and Purcell [{\em Biophys. J.} {\bf 20}, 193 (1977)] calculated how the accuracy of concentration sensing by single-celled organisms is limited by noise from the small number of counted molecules. 
Here we generalize their results to the sensing of concentration ramps, which is often the biologically relevant situation ({\em e.g.} during bacterial chemotaxis). We calculate lower bounds on the uncertainty of ramp sensing by three measurement devices: a single receptor, an absorbing sphere, and a monitoring sphere. We contrast two strategies, simple linear regression of the input signal versus maximum likelihood estimation, and show that the latter can be twice as accurate as the former. Finally, we consider biological implementations of these two strategies, and identify possible signatures that maximum likelihood estimation is implemented by real biological systems.

\end{abstract}

\maketitle


Cells are able to sense concentration gradients with high accuracy. Large eukaryotic cells such as the amoeba {\em Dictyostelium discoideum} and the budding yeast {\em Saccharomyces cerevisiae} can sense very shallow spatial gradients by comparing concentrations across their lengths \cite{Arkowitz:1999p6171}.
By contrast, small motile bacteria such as {\em Escherichia coli} detect spatial gradients indirectly by measuring concentration ramps (temporal concentration changes) as they swim \cite{Macnab:1972p5864}, and can respond to concentrations as low as 3.2 nM---about three molecules per cell volume \cite{Mao:2003p5862}. The noise arising from the small number of detected molecules sets a fundamental physical limit on the accuracy of concentration sensing, as originally shown in the seminal work of Berg and Purcell \cite{Berg:1977p3458,Bialek:2005p5858}. This approach was recently extended to derive a fundamental bound on the accuracy of direct spatial gradient sensing \cite{Endres:2008p925}.
However, no theory exists for the physical limit of ramp sensing, which is what bacteria actually do when they chemotact. In this Letter, we present such a theory for different measurement devices, from a single receptor to an entire cell.
We contrast two strategies: linear regression (LR) of the input signal (in line with Berg and Purcell) and maximum likelihood estimation (MLE) \cite{likelihood,Endres:2009p3438}, a method from statistics to optimally fit a model to data, revealing an up to twofold advantage for the latter.
Finally, we introduce a biochemical signaling network, similar to the {\em E. coli} chemotaxis system, that outputs an estimate of the ramp rate. Consistent with the derived theoretical bounds, we find that a mechanism emulating MLE yields twofold higher accuracy that one emulating LR. However, this improved performance has a cost: either storage of signaling proteins near the receptors, or irreversibility of the receptor cycle with concomitant energy consumption.



\begin{figure}
\includegraphics[width=\linewidth]{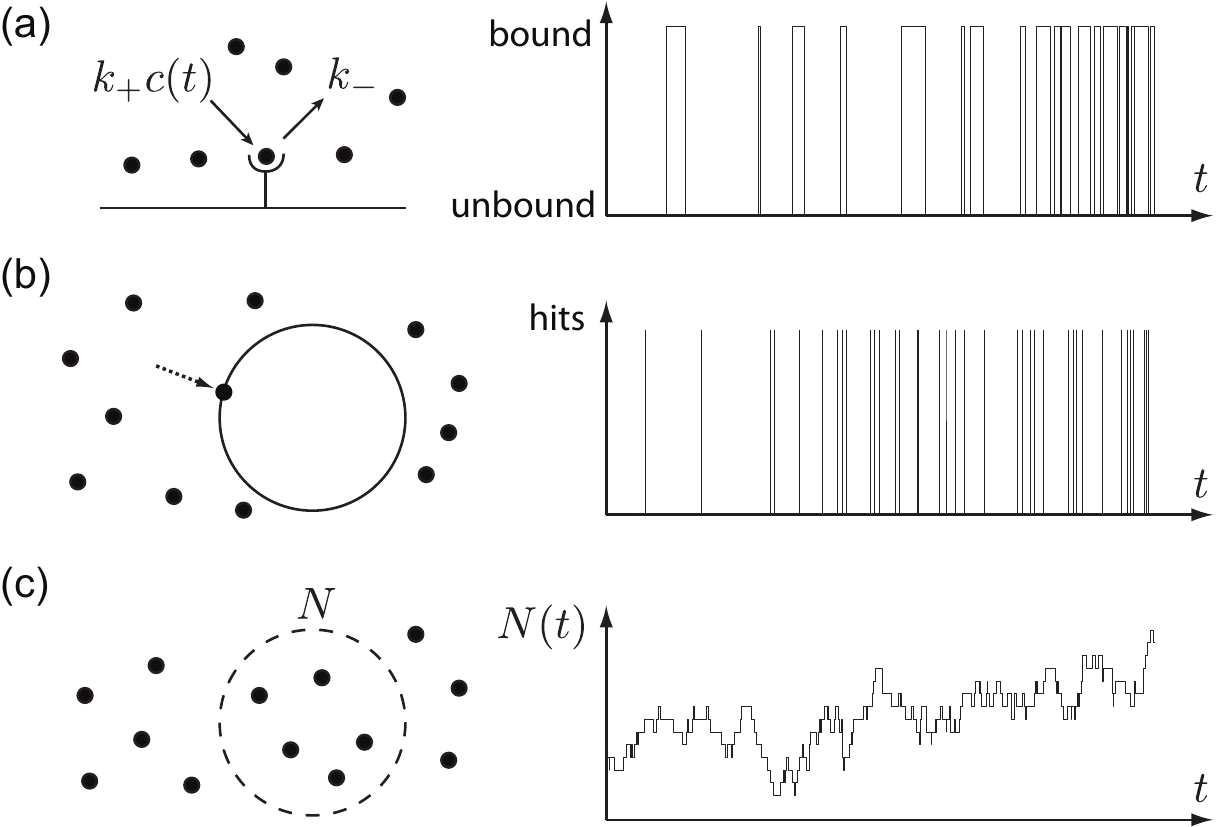}
\caption{Schematic of measurement devices and corresponding time traces for linearly increasing concentration $c(t)=c_0+c_1t$. (a) Left: a single receptor binds a particle at rate $k_+c(t)$, and releases it at rate $k_-$. Right: binary time series of receptor occupancy. (b) Left: particles are incident on an absorbing sphere with average flux $4\pi Dac(t)$. Right: sequence of times when a particle hits the sphere. (c) Left: a monitoring sphere counts the number of particles inside its volume without hindering their diffusion. Right: number $N(t)$ of particles inside the sphere as a function of time.
\label{fig:measurers}
}
\end{figure}

Sensing small numbers of molecules implies relative noise ${\sim}n^{-1/2}$, where $n$ is the number of detected molecules.
Berg and Purcell (BP) calculated how this noise affects the accuracy of concentration sensing \cite{Berg:1977p3458}. They considered three types of measurement devices: a single receptor, a perfectly absorbing sphere, and a perfectly monitoring sphere. Following their approach,
we investigate ramp sensing by these three devices when presented with a concentration $c(t)=c_0+c_1t$, as schematized in Fig.~\ref{fig:measurers}.

A single receptor [Fig.~\ref{fig:measurers}(a)] binds particles at rate $k_+c(t)$ and unbinds them at rate $k_-$. Following BP, we assume that diffusion is fast enough that the receptor never rebinds the same particle.
An ideal observer has access to the binary time series $s(t)$ of receptor occupancy between $-T/2$ and $T/2$.
The lengths of bound and unbound invervals have exponential distributions with means $1/k_-$ and $1/k_+c$, respectively. Throughout, we assume that the ramp is shallow, $c_1T\ll c_0$, and that the observation time is long compared to receptor kinetics, $T\gg 1/k_-, 1/k_+c$.
In BP, the true concentration $c$ is estimated from the fraction of time the receptor is bound, $\bar{s}=\frac{1}{T}\int_{-T/2}^{T/2} dt\, s(t)$, which is equal to the equilibrium occupancy in the limit of large times:
\beq\label{occupancy}
\bar{s}\approx \<s\>=k_+c/(k_-+k_+c),
\eeq
where $\<\cdot\>$ represents an ensemble average.
Following a similar strategy, we can estimate the ramp rate by performing the linear regression of $s(t)$ to $s_0+s_1t$:
\beq
s_0= \frac{1}{T}\int_{-\frac{T}{2}}^{\frac{T}{2}} dt\,s(t),\quad
s_1= \frac{12}{T^3}\int_{-\frac{T}{2}}^{\frac{T}{2}} dt\,t\, s(t),
\eeq
from which the concentration and the ramp rate are estimated using \eqref{occupancy} as:
\beq
c_0^{\rm LR}:=\frac{k_-}{k_+}\frac{s_0}{1-s_0},\quad
c_1^{\rm LR}:=c_0^{\rm LR}\frac{s_1}{s_0(1-s_0)}.
\eeq
The uncertainties of these estimates can be calculated from the time correlations of receptor occupancy (see Appendix \ref{appA1a}), yielding:
\bea
\frac{\<(\delta c_0^{\rm LR})^2\>}{c_0^2}=\frac{2}{n},\quad
\frac{\<(\delta c_1^{\rm LR})^2\>}{(c_0/T)^2}=\frac{24}{n},\label{unlr}
\eea
where $n$ is the total number of binding events in the time $T$. Note that the result for $c_0$ is precisely that of BP \cite{Berg:1977p3458,Endres:2009p3438}.

In \cite{Endres:2009p3438}, it was shown that the accuracy of concentration sensing could be improved using maximum likelihood estimation.
In this scheme, the parameters of the model are chosen to maximize the probability (``likelihood'') that the observed data was generated by the model. Can we also improve the accuracy of ramp sensing over LR by using this method? The time trace $s(t)$ can be characterized by the series of binding $(t_i^+)$ and subsequent unbinding $(t_i^-)$ times, $i=1,\ldots,n$. The probability of the data within our model is \cite{Endres:2009p3438}:
\beq\label{likelihood}
P=e^{-k_-T_b}e^{-k_+\sum_{i} \int_{t_i^-}^{t_{i+1}^+} dt c(t) } k_-^{n}\prod_{i=1}^n k_+c(t_i^+),
\eeq
where $T_b$ is the total bound time.
The concentration and the ramp rate, $c_0$ and $c_1$, are the model parameters.
Given the times of the events, the likelihood is maximized with respect to $c_0$ and $c_1$ by solving
${\partial P}/{\partial c_0}=0$ and ${\partial P}/{\partial c_1}=0$,
from which the maximum likelihood estimate $(c_0^{\rm MLE},c_1^{\rm MLE})$ is obtained.
In general these equations have no simple solution, but we can obtain the average behavior by exploiting the fact that binding and unbinding are fast with respect to concentration changes, {\em i.e.} that the receptor remains adiabatically in equilibrium with the concentration $c(t)$. We can thus simplify the sum and product in \eqref{likelihood}:
\bea
\sum_{i=1}^n \int_{t_i^-}^{t_{i+1}^+} dt c(t)&\approx& \int_{-\frac{T}{2}}^{\frac{T}{2}} dt\, [1-\<s(t)\>] c(t), \\
\sum_{i=1}^n \log c(t_i^+)&\approx& \int_{-\frac{T}{2}}^{\frac{T}{2}} dt\,k_- \<s(t)\>\log c(t),
\eea
where $\<s(t)\>$ is the equilibrium occupancy at time $t$, given by \eqref{occupancy} with $c=\tilde c_0+\tilde c_1 t$, where $\tilde c_0$ and $\tilde c_1$ are the {\em true} parameters that generated the data.
Applying this approximation to ${\partial P}/{\partial c_0}$, ${\partial P}/{\partial c_1}$, we confirm that $c_0^{\rm MLE}=\tilde c_0$ and $c_1^{\rm MLE}=\tilde c_1$ for $T\to\infty$ (see Appendix \ref{appA1b}). For finite times, the errors in $c_0^{\rm MLE},c_1^{\rm MLE}$ can be estimated by the Cram\'er-Rao bound \cite{Cramer}, which states that the variance of parameter estimates exceeds the inverse of the Fisher information, and approaches equality in the limit of long time series:
\beq\label{rao}
\<\delta \mathbf{c}^{\rm T}\delta \mathbf{c}\> \gtrsim -\left[\partial_{\mathbf{c}}^{\rm T}\partial_{\mathbf{c}} \log P\right]^{-1},
\eeq
where $\delta \mathbf{c} = (c_0^{\rm MLE}-\tilde c_0, c_1^{\rm MLE}-\tilde c_1)$ and $\partial_\mathbf{c} = (\partial/\partial c_0, \partial/\partial c_1)$.
Again we can use the adiabatic approximation to compute the Hessian of the log-likelihood on the right-hand side of \eqref{rao}, to obtain:
\bea
\frac{\<(\delta c_0^{\rm MLE})^2\>}{c_0^2} = \frac{1}{n},\quad
\frac{\<(\delta c_1^{\rm MLE})^2\>}{(c_0/T)^2} = \frac{12}{n}.\label{unml}
\eea
These variances are half the ones obtained from LR \eqref{unlr}. 
The first result for constant concentrations is that of \cite{Endres:2009p3438}. 
As observed there, the LR estimate adds the uncertainties from both bound and unbound interval durations.
In contrast, the maximimum likelihood estimate relies only on unbound interval durations, since these carry all the information about the concentration.

We now turn to ramp sensing by an entire cell, starting with the case of an idealized absorbing sphere [Fig.~\ref{fig:measurers}(b)]. An ideal observer witnesses a time series of absorption events, described by the instantaneous current $I(t)=\sum_{i=1}^n \delta(t-t_i)$, where $\delta(t)$ is the Dirac delta function and $\{t_i\}$ are the absorption times. The average current of molecules impinging on the sphere is given by $\<I(t)\>=4\pi Dac(t)$, where $D$ is the diffusivity, $a$ the sphere radius and $c(t)$ the concentration far from the sphere  \cite{Berg:1977p3458}.
Applying the same methods used for the single receptor, we calculated the uncertainty of ramp sensing for linear regression of $I(t)$ as well as for MLE (see Appendix \ref{appA2}). We found no difference between the two strategies, which both yield the
same uncertainties as in \eqref{unml}, with $n$ now the total number of molecules absorbed during time $T$: $n\approx 4\pi Dac_0T$.
For a monitoring sphere [Fig.~\ref{fig:measurers}(c)], molecules are free to diffuse into and out of the sphere, and the observer records the number $N(t)$ of particles inside the sphere as a function of time. On average this number is $\<N(t)\>=(4/3)\pi a^3 c(t)$. Performing a linear regression of $N(t)$ to $N_0+N_1t$, one can estimate the concentration and the ramp rate through $c_0^{\rm LR}:=3 N_0/4\pi a^3$ and $c_1^{\rm LR}:=3 N_1 /4\pi a^3$. Following \cite{Berg:1977p3458}, the uncertainty of these estimates can be calculated from the time autocorrelation of $N(t)$ (see Appendix \ref{appA3}), yielding:
\beq\label{conmon}
\frac{\<(\delta c_0)^2\>}{c_0^2}=\frac{3}{5\pi Da c_0 T},\quad
\frac{\<(\delta c_1)^2\>}{(c_0/T)^2}=\frac{36}{5\pi Da c_0 T}.
\eeq
The first result was obtained in \cite{Berg:1977p3458}. Maximum likelihood is difficult to implement in the context of the monitoring sphere because it requires a sum over all possible histories of particles exiting and returning to the sphere. Thus, whether the LR result can be improved upon remains an open question.


\begin{figure}
\includegraphics[width=\linewidth]{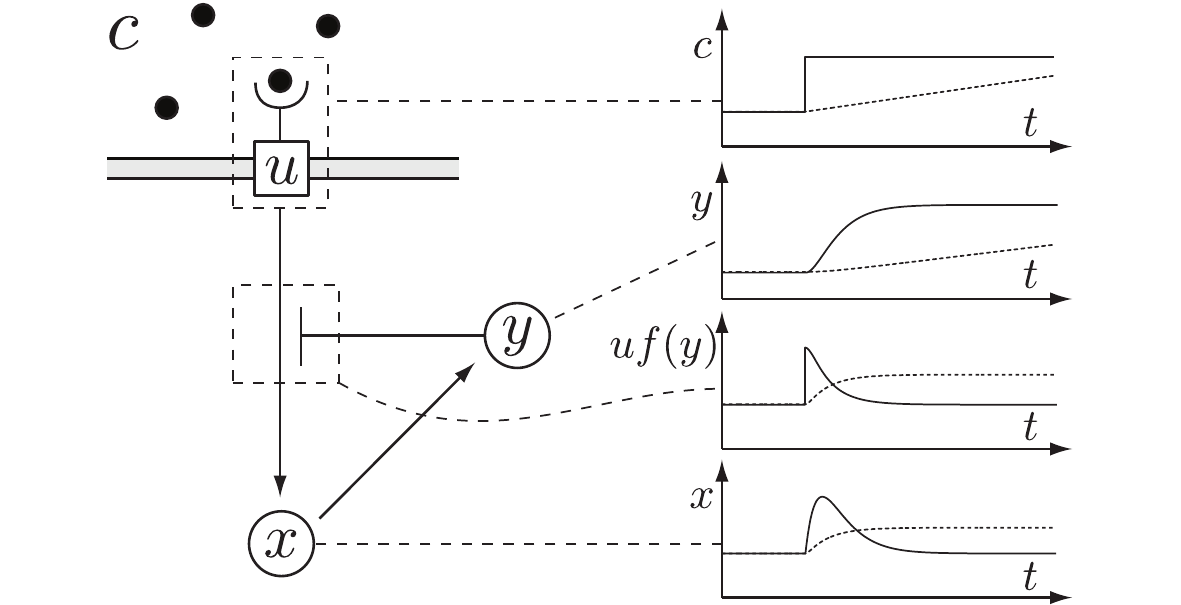}
\caption{Biochemical network for measuring concentration ramps.
Binding of ligand to the receptor increases its activity $u$ and causes species $x$ to be produced. This production is downregulated by a feedback factor $y$ which is itself catalyzed by $x$.
Right: average network response to a step function in the concentration, $c(t)=c_0+\Delta c\,\theta(t-t_0)$ (solid curves) and to a ramp, $c(t)=c_0+c_1(t-t_0)\theta(t-t_0)$ (dotted curves). In response to the step function, the network adapts precisely and $x$ decays back to its original value after an initial increase. In response to a ramp, $x$ shifts by an amount proportional to the ramp rate. The quantitative ability of the network to sense such ramps depends on whether receptors signal continuously or in a discrete burst upon ligand binding.
\label{fig:networks}
}
\end{figure}


Maximum likelihood estimation is in general the optimal way to sense ramps, and provides a twofold improvement over simple linear regression in the case of the single receptor. Could MLE be implemented in biological systems? To address this question, we now introduce a simple, deterministic biochemical network (Fig.~\ref{fig:networks}) that can approach the optimal performance limit set by MLE. The same network implements either LR or MLE depending on the receptor signaling mechanism: LR is implemented if each receptor signals continuously while a particle is bound; MLE is implemented if each receptor signals with a fixed-size burst upon binding a particle, and then releases the particle rapidly. The first case corresponds to integrating the fraction of time the receptor is bound, while the second corresponds to counting binding events. 
Accordingly, we will show that the shot noise (Poisson noise) due to the stochastic nature of binding and unbinding is twice as large in the first case as in the second.
Let $u(t)$ be the receptor activity, proportional to the instantaneous production rate of signaling molecules. For continuous signaling, this activity is simply proportional to receptor occupancy: $u(t)=\alpha s(t)$, whereas for burst signaling, $u(t)$ is a series of fixed-size bursts at the times of binding: $u(t)=\beta \sum_{i}^{n}\delta(t-t_i^+)$. 
Without loss of generality, we set $\alpha=k_-$ and $\beta=1$ so that $\<u(t)\>$ is equal to the mean rate of binding events in both cases, $\<u(t)\>=k_-k_+c(t)/(k_-+k_+c(t))$.
For averaging times much longer than $1/k_-$ and $1/k_+c$, we can approximate the fluctuations of $u(t)$ by Gaussian white noise, $u(t)=\<u(t)\>+\delta u(t)$, where $\<\delta u(t)\delta u(t')\>=g\<u(t)\>\delta(t-t')/[1+k_+c(t)/k_-]^2$,
with $g=2$ for continuous signaling, and $g=1+(k_+c/k_-)^2$ for fixed-size burst signaling (see Appendix \ref{appB1}). For rapid unbinding, $k_-\to+\infty$,
we recover the same twofold difference as between \eqref{unlr} and \eqref{unml}, and for the same reason: in the case of continuous signaling, noise from the stochasticity of bound intervals adds to the noise from random arrivals.

To extract the ramp rate from receptor activity requires a network that ``takes the derivative'' of its input signal.
An example is the {\em E. coli} chemotaxis system, which relies on precise adaptation via integral feedback \cite{Barkai:1997p906,Yi:2000p2861}. A minimal deterministic version of such a network is schematized in Fig.~\ref{fig:networks} and described by the following differential equations:
\beq
\frac{dx}{dt}=k_x\left[uf(y)-x\right],\quad
\frac{dy}{dt}=k_y(x-1)\label{net},
\eeq
where for simplicity $u(t)$ is the activity of a single receptor, and $x$ is the concentration of signaling molecules it produces.
$f(y)$ is a monotonically decreasing function regulating the production of $x$. The role of $y$ is similar to that of the receptor methylation level in {\em E. coli}: $y$ precisely adapts the production rate of signaling molecules so that the steady-state value of $x$ does not depend on the external ligand concentration.
This property is illustrated by the graphs on the right side of Fig.~\ref{fig:networks}, which show how the network responds to a sudden change in ligand concentration (solid curves). While the network output $x$ is insensitive to the absolute concentration, it responds to steady ramps (dotted curves). When the input varies slowly in time, $\<u(t)\>=u_0+u_1t$ (with $u_1\ll u_0k_x,u_0k_y$), the system responds by shifting $x$ away from $1$ so that the change in $y(t)$ tracks the change in $u(t)$:
\beq\label{readout}
\<x(t)\>=1+\gamma\frac{u_1}{k_yu_0},\
\<y(t)\>=y_0-\gamma^2\frac{u_1}{k_yu_0}-\gamma\frac{u_1}{u_0}t,
\eeq
with $u_0f(y_0)=1$
and $\gamma=-{f(y_0)}/{f'(y_0)}$.
Thus $y$ provides a readout of the absolute concentration, and $x$ provides a readout of the ramp rate. The accuracy of these representations is limited by the ligand binding shot noise $\delta u(t)$. The effect of noise can be calculated by expanding the solution of \eqref{net} linearly around its average (see Appendix \ref{appB2}):
\beq
\left[\begin{array}{c} \delta x(t)\\\delta y(t)\end{array}\right]
:=\left[\begin{array}{c}x(t)-\<x(t)\>\\y(t)-\<y(t)\>\end{array}\right]=\int_{-\infty}^tdt'\, \mathbf{K}(t-t')\delta u(t')
\nonumber
\eeq
\beq
\textrm{with }\mathbf{K}(t)=\frac{k_x}{u_0}e^{-k_xt/2}
\left[\begin{array}{c}
\cosh(\omega t)-\frac{k_x}{2\omega}\sinh (\omega t)\\
\frac{k_y}{\omega}\sinh (\omega t)
\end{array}\right],\nonumber
\eeq
where $\omega^2=k_x^2/4-k_xk_y/\gamma$ ($\omega$ can be imaginary).
From \eqref{readout} we deduce the uncertainties of $c_0$ and $c_1$:
\beq
\frac{\<(\delta c_0)^2\>}{c_0^2}=\frac{gk_y/\gamma}{2u_0},\qquad
\frac{\<(\delta c_1)^2\>}{(c_0k_y/\gamma)^2}=\frac{gk_x}{2u_0}.
\eeq
For a fixed $k_y$, the optimal value of $k_x$ is the smallest one with a non-oscillating response kernel $\mathbf{K}(t)$: $k_x=4k_y/\gamma$. Systems with oscillating kernels are undesirable because they detect oscillations rather than ramps. For $k_x=4k_y/\gamma$,
our results are consistent with those of Eqs.~\eqref{unlr} and \eqref{unml}, namely uncertainties inversely proportional to the number of binding events, if we interpret $\gamma/k_y \to T$ as the effective time of measurement, and $u_0$ as the rate of binding events.
The factor $g$ reflects the difference between the two mechanisms of receptor signaling.

Despite its simplicity, our biochemical model may help analyze features of real biological systems.
There are two separate aspects to the model:
on the input side, different mechanisms of receptor signaling---continuous signaling (LR) versus burst signaling (MLE)---affect readout accuracy; on the output side, integral feedback provides a natural readout for sensing ramps.

Many receptors, including the well-studied chemotaxis receptors of {\em E. coli}, signal continuously rather than in bursts, and therefore do not employ MLE. In practice, how could cells implement MLE?
A receptor could simply ``store'' a fixed amount of signaling molecules and release all of them upon ligand binding. Alternatively, receptors could signal continuously following a binding event but with a narrowly peaked distribution of durations.
Our results can easily be extended to an arbitrary distribution of durations $\tau_b$, yielding
$g=1+\<(\delta\tau_b)^2\>/\<\tau_b\>^2$:
the more peaked the distribution of $\tau_b$, the less noisy the readout. For equilibrium binding/unbinding, we find $g\geq 2$ (see Appendix \ref{appC}), with an irreversible binding cycle driven by energy dissipation required to achieve $g<2$.
Interestingly, there are examples of such irreversible cycles in ligand-gated ion channels \cite{Schneggenburger:1997p6174,Wang:2002p6173,Csanady:2009p6215}, where ions play the role of our output signal $x$.
In these ion channels, peaked open-time distributions are interpreted as evidence that time reversibility is broken and energy is being consumed \cite{Colquhoun}. We speculate that the role of this irreversibility may be to reduce the variance of bursts, thereby increasing the accuracy of concentration or ramp sensing.
Relatedly, a multiplicity of irreversible steps in rhodopsin signaling has been shown to explain the reproducibility of single-photon responses in rod cells \cite{Doan:2006p6626}.

As for the mechanism of ramp sensing, the integral feedback system underlying {\em E. coli} chemotaxis is similar to our simple model. However, the receptor methylation level, which plays the same role as $y$ in our model, adjusts the binding/unbinding rates $k_+/k_-$ so that $k_-\approx k_+c$, rather than adjusting the production rate $k_xf(y)u$ as in \eqref{net}. 
In {\em E. coli} receptors increase their gain by responding cooperatively \cite{Endres:2006p945}, and $ k_-\approx k_+c$ is required to maximize this gain, which precludes the limit $k_-\ll k_+c$ required for MLE. Moreover, $k_+$ is physically limited by diffusion and receptor size, and should optimally be kept near the diffusion limit to maximize the number of binding events.
It is worth noting that in {\em E. coli} the methylation and demethylation processes responsible for integral feedback are themselves subject to noise, giving rise to additional fluctuations \cite{Emonet:2008p2793}.
For a receptor signaling in bursts, integral feedback could act by adjusting the number of released molecules upon binding if the receptor stores molecules, or the mean bound duration $\<\tau_b\>$ if signaling is continous, or the channel conductivity in ligand-gated ion channels.
We hope that our analysis will suggest experiments for testing these scenarios.

We thank Pankaj Mehta and Aleksandra Walczak for helpful suggestions.
T. M. was supported by the Human Frontier Science Program and N.S.W. by National Institutes of Health Grant No. R01 GM082938.


\appendix

\section{Uncertainties in ramp sensing}\label{appA}

\subsection{Single receptor} \label{appA1}
We consider a single receptor, which particles bind to and unbind from stochastically. The unbinding rate is denoted by $k_-$, and the binding rate by $k_+c(t)$, where $c(t)$ is the concentration of particles.
We assume fast diffusion and so neglect rebinding of particles \cite{Bialek:2005p5858}.
Between the times $-T/2$ and $T/2$, the concentration changes slowly with time: $c(t)=c_0+c_1t$. We assume that variations are small: $c_1T\ll c_0$, and that the integration time $T$ is large compared to the waiting times: $T\gg k_-^{-1}, (k_+ c_0)^{-1}$.

\subsubsection{Linear regression} \label{appA1a}
We perform linear regression of the binary time series $s(t)$ of receptor occupancy $s(t)$ to $s_0+s_1t$.
The time trace $s(t)$ can be characterized by the series of binding $(t_i^+)$ and subsequent unbinding $(t_i^-)$ times, $i=1,\ldots,n$. 
The resulting estimates for $c_0$ and $c_1$ are:
\bea
c_0^{\rm LR}&=&\frac{k_-}{k_+}\frac{s_0}{1-s_0}\label{c0_1},\\
c_1^{\rm LR}&=&c_0^{\rm LR}\frac{s_1}{s_0(1-s_0)}\label{c1_1},
\eea
where
\bea
s_0&=& \frac{1}{T}\int_{-T/2}^{T/2} dt\,s(t) = \frac{1}{T}\sum_i (t_{i+1}^--t_i^+),\\
s_1&=& \frac{12}{T^3}\int_{-T/2}^{T/2} dt\,t\, s(t)=\frac{6}{T^3}\sum_i ({t_{i+1}^{-}}^2-{t_i^{+}}^2),
\eea
are the results of the linear regression.

In the limit of long time series, these estimates give the correct answers: replacing $s(t)$ by its ensemble average value, $\< s(t)\>=k_+(\tilde c_0+\tilde c_1t)/[k_-+k_+(\tilde c_0+\tilde c_1t)]$, where $\tilde c_0$,$\tilde c_1$ are the true values of the concentration and the ramp rate, one obtains:
\bea
\<s_0\>&=&\frac{k_+\tilde c_0}{k_-+k_+\tilde c_0},\\
\<s_1\>&=&\frac{k_-k_+\tilde c_1}{(k_-+k_+c_0)^2},
\eea
which leads to $c_0^{\rm LR}=\tilde c_0$ and $c_1^{\rm LR}=\tilde c_1$ on average. The expected error can be obtained from the covariance matrix $\<\delta \mathbf{s}{}^{t}\delta \mathbf{s}\>$.
To compute this covariance matrix we need the following quantity:
\beq
\<s(t)s(t+t')\>-\<s(t)\>\<s(t+t')\>\approx \frac{k_-k_+c(t) e^{-|t'|[k_-+k_+c(t)]}}{[k_-+k_+c(t)]^2},
\eeq
where the approximation $\approx$ is valid provided that $|t'|\ll c_0/c_1$. In the limit of large times we have:
\bea
\<(\delta s_0)^2\>&\approx&\frac{2}{T}\frac{k_-k_+c_0}{(k_-+k_+c_0)^3},\\
\<(\delta s_1)^2\>&\approx&\frac{24}{T^3}\frac{k_-k_+c_0}{(k_-+k_+c_0)^3},\\
\<\delta s_0\delta s_1\>&\approx &0,
\eea
from which we deduce (using Eqs.~\ref{c0_1}, \ref{c1_1} and $c_1T\ll c_0$):
\bea
\frac{\<(\delta c^{\rm LR}_0)^2\>}{c_0^2}&=&\frac{2}{n},\label{unun2}\\
\frac{\<(\delta c^{\rm LR}_1)^2\>}{(c_0/T)^2}&=&\frac{24}{n}.\label{ungr2}
\eea
The above result for $c_0$ was originally obtained by Berg and Purcell \cite{Berg:1977p3458}.

\subsubsection{Maximum likelihood}\label{appA1b}

The probability (``likelihood'') of a sequence of binding and unbinding events at times $\{t_i^+\}$ and $\{t_i^-\}$ between $-T/2$ and $T/2$, is:
\beq\label{logL0}
P=e^{-k_-T_b}e^{-k_+\sum_{i=1}^n \int_{t_i^-}^{t_{i+1}^+} dt c(t) } {(k_-k_+)}^{n}\prod_{i=1}^nc(t_i^+).
\eeq
The log-likelihood is (up to a constant independent of $c_0$ and $c_1$):
\beq\label{logL}
\begin{split}
\log P=\textrm{const}-k_+\sum_i \left (c_0+c_1\frac{t_{i+1}^++t_i^-}{2}\right)(t_{i+1}^+-t_i^-) \\
+ \sum_i \log(c_0+c_1t_i^+).
\end{split}
\eeq
Given the times of the binding and unbinding events, the optimal strategy for estimating the concentration and the ramp rate is to maximize this log-likelihood with respect to $c_0$ and $c_1$, {\em i.e.} to solve
\beq\label{logP}
\frac{\partial\log P}{\partial c_0}=0\quad\textrm{and}\quad \frac{\partial\log P}{\partial c_1}=0,
\eeq
from which we can obtain the maximum likelihood estimates $c_0^{\rm MLE}$ and $c_1^{\rm MLE}$.
To solve Eq.~\ref{logP}, we exploit the self-averaging property of $\log P$ as $T\to +\infty$ (adiabatic approximation). The sums in Eq.~\ref{logL0} and \ref{logL} become:
\bea
\sum_{i=1}^n \int_{t_i^-}^{t_{i+1}^+} dt c(t)&\approx& \int_{-\frac{T}{2}}^{\frac{T}{2}} dt\, [1-\<s(t)\>] c(t), \label{rep1}\\
\sum_{i=1}^n \log c(t_i^+)&\approx& \int_{-\frac{T}{2}}^{\frac{T}{2}} dt\,k_- \<s(t)\>\log c(t),\label{rep2}
\eea
where $\<s(t)\>$ is the equilibrium probability of the receptor being bound at time $t$:
\beq\label{occupancy}
\<s(t)\>=\frac{k_+\tilde c(t)}{k_-+k_+\tilde c(t)},
\eeq
and $\tilde c(t)=\tilde c_0+\tilde c_1 t$ is the true concentration.
We take the derivatives of the log-likelihood (Eq.~\ref{logL}, with replacements from Eqs.~\ref{rep1} and \ref{rep2}) with respect to $c_0$ and $c_1$:
\bea
\frac{\partial\log P}{\partial c_0}&=&-\int_{-T/2}^{T/2} dt\, \frac{k_-k_+}{k_-+k_+(\tilde c_0+\tilde c_1 t)} \\ \nonumber &&+ \int_{-T/2}^{T/2} dt\,\frac{k_-k_+(\tilde c_0+\tilde c_1t)}{[k_-+k_+(\tilde c_0+\tilde c_1 t)](c_0+c_1 t)},\\
\frac{\partial\log P}{\partial c_1}&=&-\int_{-T/2}^{T/2} dt\, \frac{k_-k_+ t}{k_-+k_+(\tilde c_0+\tilde c_1 t)} \\ \nonumber &&+ \int_{-T/2}^{T/2} dt\,\frac{k_-k_+(\tilde c_0+\tilde c_1t)t}{[k_-+k_+(\tilde c_0+\tilde c_1 t)](c_0+c_1 t)}.
\eea
It is straightforward to confirm that these expressions become zero for $c_0=\tilde c_0$ and $c_1=\tilde c_1$. 
Thus, in the limit of long time series, the maximum likelihood estimates coincide with the true values of the concentration and the ramp rate. The expected error of these estimates is given by the Cram\'er-Rao inequality, which becomes an equality in the limit of large $T$:
\beq\label{rao}
\<\delta \mathbf{c}^{\rm T}\delta \mathbf{c}\> \gtrsim -\left[\partial_{\mathbf{c}}^{\rm T}\partial_{\mathbf{c}} \log P\right]^{-1},
\eeq
where $\delta \mathbf{c} = (c_0^{\rm MLE}-\tilde c_0, c_1^{\rm MLE}-\tilde c_1)$ and $\partial_\mathbf{c} = (\partial/\partial c_0, \partial/\partial c_1)$.

The second derivatives of the log-likelihood are, to leading order:
\bea
\frac{\partial^2\log P}{\partial c_0^2}&=& -\int_{-T/2}^{T/2} dt\,\frac{k_-k_+(\tilde c_0+\tilde c_1t)}{[k_-+k_+(\tilde c_0+\tilde c_1 t)](c_0+c_1 t)^2}\nonumber\\&\approx & -\frac{T}{c_0} \frac{k_-k_+}{k_-+k_+c_0},\label{eq1}\\
\frac{\partial^2\log P}{\partial c_1^2}&=& -\int_{-T/2}^{T/2} dt\,\frac{k_-k_+(\tilde c_0+\tilde c_1t)t^2}{[k_-+k_+(\tilde c_0+\tilde c_1 t)](c_0+c_1 t)^2}\nonumber\\&\approx & -\frac{T^3}{12c_0} \frac{k_-k_+}{k_-+k_+c_0},\label{eq2}\\
\frac{\partial^2\log P}{\partial c_0\partial c_1}&=& -\int_{-T/2}^{T/2} dt\,\frac{k_-k_+(\tilde c_0+\tilde c_1t)t}{[k_-+k_+(\tilde c_0+\tilde c_1 t)](c_0+c_1 t)^2}\nonumber\\&\approx & 0.\label{eq3}
\eea
Exploiting the diagonal structure of the Hessian, we obtain:
\bea
\frac{\<(\delta c^{\rm MLE}_0)^2\>}{c_0^2} &=& \frac{1}{n},\label{unun1}\\
\frac{\<(\delta c^{\rm MLE}_1)^2\>}{(c_0/T)^2} &=& \frac{12}{n},\label{ungr1}
\eea
where $n$ is the total number of binding events in the time series, $n \approx k_-k_+c_0T/(k_-+k_+c_0)$. The above result for $c_0$ was first obtained in \cite{Endres:2008p925}.

\subsection{Absorbing sphere}\label{appA2}

We now consider as a measurement device a perfectly absorbing sphere of radius $a$.
Solving Laplace's equation with absorbing boundary conditions at the surface of the sphere yields the average flux of particles impinging on the sphere \cite{Berg:1977p3458}:
\beq
\<I(t)\> =4\pi Da c(t),
\eeq
where $D$ is the diffusivity, $a$ is the sphere radius, and $c(t)=c_0+c_1t$ is the (time-dependent) concentration far away from the sphere. When the ramp arises from motion of the sphere up a spatial gradient (with velocity $v$), one must be careful: in general the flux of particles is not uniform around the sphere, and will differ between the front and the rear of the sphere (windshield effect) \cite{Berg:1977p3458}. However, this effect can be neglected when measuring the change of concentration with time if the particle turnover rate $D/a^2$ is much larger than rate of moving one body length $v/a$, {\em i.e.} if $D/a\gg v$. For bacteria the relevant numbers are $D \approx 10^3$
$\mu$m${}^2$/s, $a \approx$ 1 $\mu$m, and $v \approx 30$ $\mu$m/s, so that $D/a \approx 10^3$ $\mu$m/s. Therefore, from the perspective of a swimming bacterium, the concentration appears to be changing with time, essentially uniformly in space, which is the case we consider.

\subsubsection{Linear regression}
For a constant concentration, the simplest estimate for the concentration is $c^{\rm LR}_0=n/4\pi DaT$, where $n$ is the total number of absorption events in time $T$.
In the presence of a ramp, an estimate of both concentration and ramp rate can be obtained by performing the linear regression of
\beq
I(t)=\sum_i \delta(t-t_i)
\eeq
to $4\pi Da (c_0+c_1t)$, where the $\{t_i\}$ are the absorption times, yielding:
\beq
\begin{split}
&\frac{\partial}{\partial c_0}\int_{-T/2}^{T/2}dt\,{\left[\sum_{i=1}^n \delta(t-t_i)-4\pi Da (c_0+c_1t)\right]}^2\\ &\Longrightarrow
c_0^{\rm LR}=\frac{n}{4\pi DaT},
\end{split}
\eeq
\beq
\begin{split}
&\frac{\partial}{\partial c_1}\int_{-T/2}^{T/2}dt\,{\left[\sum_{i=1}^n \delta(t-t_i)-4\pi Da (c_0+c_1t)\right]}^2\\ &\Longrightarrow
c_1^{\rm LR}=\frac{12\sum_{i=1}^n t_i}{4\pi DaT^3}.
\end{split}
\eeq
Since the particle absorptions are independent events, $n$ is a Poisson variable, $\<(\delta n)^2\>=\<n\>=4\pi Da\tilde c_0$ (as above, $\tilde c_0,\tilde c_1$ are the true values of the concentration and the ramp rate), and thus $\<(\delta c_0)^2\>/c_0^2=1/n$.
Concerning the ramp rate, we have
\beq
\left\<\sum_{i=1}^n t_i\right\>=\int_{-T/2}^{T/2} dt\,t\<I(t)\>={4\pi Da c_1} \frac{T^3}{12},
\eeq
and
\beq
\begin{split}
\left\<{\delta\left(\sum_{i=1}^n t_i\right)}^2\right\>&=\left\<{\delta \left(\int_{-T/2}^{T/2} dt\,t\delta I(t)\right)}^2\right\>\\
&=\int_{-T/2}^{T/2} dt\,t^2\<I(t)\>={4\pi Da c_0} \frac{T^3}{12},
\end{split}
\eeq
where we have used the fact that over a short time $\delta t$, the number of absorbed particles is a Poisson variable, which entails:
\beq
\left\<{\delta\left(\int_{t}^{t+\delta t}dt'\,t' \delta I(t')\right)}^2\right\>\approx \<I(t)\>t^2\delta t .
\eeq
Therefore:
\beq
\<(\delta c_1^{\rm LR})^2\>=\frac{12c_0}{4\pi Da T^3}.
\eeq
Finally, in summary:
\bea
\frac{\<(\delta c^{\rm LR}_0)^2\>}{c_0^2} &=& \frac{1}{n},\label{conabslr}\\
\frac{\<(\delta c^{\rm LR}_1)^2\>}{(c_0/T)^2} &=& \frac{12}{n}, \label{gradabslr}
\eea
where $n$ is the total number of absorption events in time $T$.

\subsubsection{Maximum likelihood}

The observations by an absorbing sphere in the interval $(-T/2,T/2)$ are summarized by the sequence of times when particles are absorbed, $\{t_i\}$, $i=1,\ldots,n$. The log-likelihood of this sequence reads:
\beq
\begin{split}
\log P(\{t_i\})&=-\int_{-T/2}^{T/2} dt\, \<I(t)\>_{c_0,c_1} + \sum_{i=1}^n \log \<I(t_i)\>_{c_0,c_1}\\
&=\textrm{const}-4\pi Da c_0 T + \sum_{i=1}^n \log (c_0+c_1t_i),
\end{split}
\eeq
where $\<I(t)\>_{c_0,c_1}$ is the expected average flux for concentration $c_0$ and ramp rate $c_1$.
Setting the derivatives to zero, $d\log P/dc_0=0$, $d\log P/dc_1=0$, yields to leading order:
\bea
{c^{\rm MLE}_0}&=&\frac{n}{4\pi Da T},\label{c0}\\
{c^{\rm MLE}_1}&=&{c^{\rm MLE}_0}\frac{\sum_i t_i}{\sum_i t_i^2}.\label{c1}
\eea
where we have used $c_1T\ll c_0$. 
In the limit of long time series, the sums self-average and become:
\beq\label{nabs}
n\approx \<n\>=4\pi D a \tilde c_0 T,
\eeq
\beq
\begin{split}
\sum_i t_i \approx \left\<\sum_i t_i \right\> & =\int_{-T/2}^{T/2} dt\, 4\pi Da (\tilde c_0+\tilde c_1 t) t \\ & = 4\pi D a \tilde c_1 \frac{T^3}{12},
\end{split}
\eeq
\beq
\begin{split}
\sum_i t_i^2 \approx \left\<\sum_i t_i^2 \right\> & =\int_{-T/2}^{T/2} dt\, 4\pi Da (\tilde c_0+\tilde c_1 t) t^2 \\ & = 4\pi D a \tilde c_0 \frac{T^3}{12},
\end{split}
\eeq
which yields $c_0^{ML}=\tilde c_0$ and $c_1^{ML}=\tilde c_1$ using Eqs.~\ref{c0} and \ref{c1}.

To obtain the uncertainties, we calculate the second derivatives of the log-likelihood and use the Cram\'er-Rao bound. After a calculation similar to Eqs.~\ref{eq1}-\ref{eq3}, we find:
\bea
\frac{\<(\delta c^{\rm MLE}_0)^2\>}{c_0^2} &=& \frac{1}{n},\label{conabs}\\
\frac{\<(\delta c^{\rm MLE}_1)^2\>}{(c_0/T)^2} &=& \frac{12}{n}. \label{gradabs}
\eea

Therefore, maximum likelihood estimation is equivalent to linear regression in the limit of long time series.
(One can similarly show that the same conclusion applies to the sensing of spatial gradients: the estimate from linear regression derived in \cite{Endres:2008p925} is equivalent to maximum likelihood estimation.)

\subsection{Monitoring sphere}\label{appA3}

\subsubsection*{Linear regression}

Following Berg and Purcell \cite{Berg:1977p3458}, we now consider a perfect monitoring sphere of radius $a$, which lets particles freely diffuse in and out of the sphere. 
This device records the number of particles $N(t)$ inside the sphere at all times, with ensemble average $\<N(t)\>=(4/3)\pi a^3 c(t)$.
The particle turnover time, $a^2/D$ (where $D$ is the diffusivity), is assumed to be small compared to the total time $T$ of observation.
Regressing $N(t)$ to $(4/3)\pi a^3 (c_0+c_1t)$ yields an estimate for the concentration and the ramp rate:
\bea
c_0^{\rm LR}&=&\frac{1}{v}\frac{1}{T}\int_{-T/2}^{T/2} dt\,N(t),\\
c_1^{\rm LR}&=&\frac{1}{v}\frac{12}{T^3}\int_{-T/2}^{T/2} dt\,N(t)\,t,
\eea
where $V=(4/3) \pi a^3$ is the volume of the sphere. In the limit of long times, this estimate gives the true value of the concentration and the ramp rate:
$\<c_0^{\rm LR}\>=\tilde c_0$, $\<c_1^{\rm LR}\>=\tilde c_1$. The uncertainties  are given by the variances of $c_0^{\rm LR}$ and $c_1^{\rm LR}$. For example, the uncertainty of the concentration estimate is:
\beq
\<(\delta c_0^{\rm LR})^2\>=\frac{1}{V^2}\frac{1}{T^2}\int_{-T/2}^{T/2} dt\,\int_{-T/2}^{T/2} dt' \,\<\delta N(t)\delta N(t')\>,
\eeq
where $\delta N(t)=N(t)-\<N(t)\>$.
To compute $\<\delta N(t)\delta N(t')\>$, we decompose $N(t)$ as a sum over $M$ independent particles in a large volume $U\gg V$: $N(t)=\sum_{i=1}^M x_i(t)$, where $x_i(t)$ is a binary variable for each particle, whose value is $1$ if the particle is in the sphere, and $0$ otherwise. We assume that $M$ and $U$ are large, such that $M/U=\tilde c_0+\tilde c_1t$. The ramp can arise from a change in volume $U$, or from creation/annihilation of particles. 
For definiteness, we will assume that the number of particles $M$ stays fixed, while the confining volume $U$ varies in time.
Since the particles are independent of each other, we have $\<\delta x_i(t)\delta x_j(t')\>=0$, and therefore:
\beq\label{tos}
\<(\delta c_0^{\rm LR})^2\>=M\frac{1}{V^2}\frac{1}{T^2}\int_{-T/2}^{T/2} dt\,\int_{-T/2}^{T/2} dt' \<\delta x(t)\delta x(t')\>,
\eeq
where $x(t)$ the binary indicator variable of a single particle, and $\delta x(t)=x(t)-\<x(t)\>$. In the limit of large time differences $|t-t'|\gg D/a^2$, the autocorrelation function $\<\delta x(t)\delta x(t')\>$ decays to $0$. Since $T\gg D/a^2$, Eq.~\ref{tos} simplifies to:
\beq
\<(\delta c_0^{\rm LR})^2\>=2M\frac{1}{V^2}\frac{1}{T}\int_{0}^{+\infty} dt \<\delta x(0)\delta x(t)\>,
\eeq
where we have used time translation invariance and time-reversal symmetry between $t$ and $t'$. The quantity $\<x(0) x(t)\>$ is the probability that a particular particle was in the sphere at times $0$ and $t$, {\em i.e.} $P(x(0)=1) P(x(t)=1|x(0)=1)=\<x(0)\> u_{\rm BP}(t)$, with $\<x(0)\>=(\tilde c_0+\tilde c_1 t)v/M$ and
\beq
\begin{split}
u_{\rm BP}(t)&:=P(x(t)=1|x(0)=1)\\
&=\frac{1}{V}\int d\vec r\,\int d\vec r' \,\frac{\exp\left[-\frac{|\vec r-\vec r'|^2}{4Dt}\right]}{(4\pi Dt)^{3/2}},
\end{split}
\eeq
where the integrals are over the sphere of radius $a$.
A calculation familiar from electrostatic done in \cite{Berg:1977p3458} yields:
\beq
\int_0^{+\infty} dt\,u_{\rm BP}(t) = \frac{2a^2}{5D}.
\eeq
Combining the above results and using $\<x(0)\>\<x(t)\>= (cV/M)^2$, we obtain the Berg and Purcell bound:
\beq\label{conmon}
\frac{\<(\delta c_0^{\rm LR})^2\>}{c_0^2}=\frac{3}{5\pi Da c_0 T}.
\eeq
The calculation of the uncertainty of the ramp rate proceeds very similarly:
\begin{widetext}
\beq
\<(\delta c_1^{\rm LR})^2\>=M\frac{1}{V^2}\frac{12^2}{T^6}\int_{-T/2}^{T/2} dt\,\int_{-T/2}^{T/2} dt'\, tt'\,\<\delta x(t)\delta x(t')\>
\approx 2M\frac{1}{V^2}\frac{12^2}{T^6}\int_{-T/2}^{T/2} dt\, t^2\,\frac{(\tilde c_0+\tilde c_1t)V}{M}\int_{0}^{+\infty}dt'\,u_{\rm BP}(t'),
\eeq
\end{widetext}
where we have used $t'\approx t$. This approximation is justified by the fact that for $|t-t'|\gg D/a^2$, $\<\delta x(t)\delta x(t')\>\to 0$.
Finally we obtain:
\beq\label{gradmon}
\frac{\<(\delta c_1)^2\>}{(c_0/T)^2}=\frac{36}{5\pi Da c_0 T}.
\eeq

\section{Uncertainties in a ramp-sensing biochemical network}\label{appB}

\subsection{Input noise}\label{appB1}
We first consider the most general case of the biochemical network shown in Fig.~2 of the main text. Ligands bind the receptor at a rate $k_+c(t)$. When a ligand binds to the receptor, it remains bound for a time $\tau_b$, and $\beta$ signaling molecules are released. The distributions of both $\tau_b$ and $\beta$ can be arbitrary.
For continuous signaling, particles are produced at a rate $\beta$ while the receptor is bound, yielding $\beta=\alpha \tau_b$, while for fixed-size burst signaling, $\beta$ particles are realeased directly upon binding at time $t_i^+$. 
Both scenarios can be described by the instantaneous receptor activity:
\bea
\textrm{Continuous signaling}& u(t)&=\alpha s(t),\nonumber\\
\textrm{Fixed-size burst signaling }& u(t)&=\beta\sum \delta(t-t_i^+),\nonumber
\eea
where, in the first equation, $s(t)=1$ when the receptor is bound and $0$ otherwise.
When there is only one bound state, $\tau_b$ has an exponential distribution, with $\<\tau_b\>=1/k_-$ and $\<(\delta\tau_b)^2\>=1/k_-^2$.
For averaging times much longer than the durations of bound and unbound intervals, $u(t)$ can be approximated by a Gaussian variable. The mean of $u(t)$ is:
\bea
\textrm{Continuous signaling} &\<u(t)\>&=\frac{\alpha k_+c(t)}{1/\<\tau_b\>+k_+c(t)},\nonumber\\
\textrm{Fixed-size burst signaling} &\<u(t)\>&=\frac{\beta}{\<\tau_b\>+1/k_+c(t)}.\nonumber
\eea
To estimate the variance of $u(t)$ for fixed-size burst signaling, we calculate the variance of the number of binding events during a time $\Delta t \gg \<\tau_b\>,1/k_+c(t)$ (but $\Delta t\ll T$). The variance of the duration of one binding/unbinding cycle is: $\<\delta (\tau_b+\tau_u)^2\>=\<(\delta\tau_b)^2\>+1/[k_+c(t)]^2$. If there are $n$ binding/unbinding events during $\Delta t$, the relative variance of $n$ for a fixed duration $\Delta t$ is equal to the relative variance of $\Delta t$ for a fixed number of binding/unbinding events $n$:
\beq
\frac{\<(\delta n)^2\>}{n^2}=\frac{\<(\delta\Delta t)^2\>}{(\Delta t)^2}=\frac{1}{n} \frac{\<(\delta\tau_b)^2\>+1/[k_+c(t)]^2}{\left[\<\tau_b\>+1/k_+c(t)\right]^2},
\eeq
hence:
\beq
\begin{split}
\left\<{\delta\left(\int_{t}^{t+\Delta t}dt'\,u(t')\right)}^2\right\>= \beta^2{\<(\delta n)^2\>}\\
=\beta^2\Delta t
\frac{\<(\delta\tau_b)^2\>+1/[k_+c(t)]^2}{\left[\<\tau_b\>+1/k_+c(t)\right]^3}.
\end{split}
\eeq
Thus, for averaging times much longer than $\<\tau_b\>$ and $1/k_+c(t)$, we may write:
\beq
\<\delta u(t)\delta u(t')\>=\beta^2 k_+c(t)\frac{1+\<(\delta\tau_b)^2\>\left[k_+c(t)\right]^2}{[1+\<\tau_b\>k_+c(t)]^3}\delta(t-t').
\eeq
Similarly, we can estimate the fluctuations of $u(t)$ for continuous signaling:
\beq
\<\delta u(t)\delta u(t')\>=\alpha^2 {\<\tau_b^2\>}\frac{k_+c(t)}{[1+\<\tau_b\>k_+c(t)]^3}\delta(t-t').
\eeq
Setting $\alpha=1/\<\tau_b\>$ and $\beta=1$ without loss of generality, we obtain for both cases:
\bea
\<u(t)\>&=&\frac{k_+c(t)}{1+\<\tau_b\>k_+c(t)},\\
\<\delta u(t)\delta u(t')\>&=& g\frac{k_+c(t)}{[1+\<\tau_b\>k_+c(t)]^3}\delta(t-t'),
\eea
with $g=1+{\<(\delta\tau_b)^2\>}/{\<\tau_b\>^2}$ for continous signaling, and $g=1+\<(\delta\tau_b)^2\>\left[k_+c(t)\right]^2$ for fixed-size burst signaling.
Note that for an exponential distribution of bound intervals durations $\<(\delta\tau_b)^2\>=\tau_b^2=1/k_-^2$, so that $g=2$ for continuous signaling, and $g=1+[k_+c(t)/k_-]^2$ for fixed-size burst signaling.
In the limit of short bound-time durations, $\tau_b\to 0$, the general results become:
\beq\label{input}
\<u(t)\>=k_+c(t),\quad\<\delta u(t)\delta u(t')\>=g\<u(t)\>\delta(t-t'),
\eeq
with $g=1+\<(\delta\tau_b)^2\>/\<\tau_b\>^2$ for continuous signaling and $g=1$ for fixed-size burst signaling.

\subsection{Output noise}\label{appB2}
Consider the network shown in Fig.~2 of the main text and described by the equations:
\bea
\frac{dx}{dt}&=&k_x\left[uf(y)-x\right],\label{net1}\\
\frac{dy}{dt}&=&k_y(x-1)\label{net2}.
\eea
When the network is presented with a slow concentration ramp $c_0+c_1t$ (such that $c_1t\ll c_0$), with the input $u(t)$ given by Eq.~\ref{input}, {\em i.e.} $u(t)=u_0+u_1t+\delta u(t)$ with:
\bea
u_0&=&k_+c_0/(1+\<\tau_b\>k_+c_0),\label{u0}\\ u_1&=&k_+c_1/(1+\<\tau_b\>k_+c_0)^2,\label{u1}\\
\<\delta u(t)\delta u(t')\>&=& \frac{g{k_+c_0}} {[1+\<\tau_b\>k_+c_0]^3}\label{du}\delta(t-t'),
\eea
the average network response is:
\bea
\<x(t)\>&=&1+\gamma\frac{u_1}{k_yu_0},\label{readout1}\\
\<y(t)\>&=&y_0-\gamma^2\frac{u_1}{k_yu_0}+\gamma\frac{u_1}{u_0}t,\label{readout2}\\
u_0f(y_0)&=&1,\label{readout3}
\eea
with $\gamma=-{f(y_0)}/{f'(y_0)}$.
We collect the fluctuations of $x$ and $y$, $\delta x(t)=x(t)-\<x(t)\>$ and $\delta y(t)=y(t)-\<y(t)\>$, into a single vector:
\beq
X=\left[\begin{array}{l} \delta x\\ \delta y\end{array}\right].
\eeq
Linearizing Eqs.~\eqref{net1} and \eqref{net2} yields:
\beq
\frac{dX}{dt}+MX=\frac{k_x}{u_0}\left[\begin{array}{c} \delta u(t)\\ 0\end{array}\right],\quad\textrm{with}\quad M=\left[\begin{array}{cc} k_x & k_x/\gamma\\ -k_y &0\end{array}\right].
\eeq
Multiplying by $e^{tM}$ on both sides and integrating, one obtains:
\beq
X=\frac{k_x}{u_0}e^{-tM}\int_{-\infty}^{t}dt'\, e^{t'M}\left[\begin{array}{c} \delta u(t)\\ 0\end{array}\right],
\eeq
with
\begin{widetext}
\beq
e^{tM}=e^{k_xt/2}
\left[\begin{array}{cc}
\cosh(\omega t)+\frac{k_x}{2\omega}\sinh(\omega t) &
\frac{k_x}{\gamma\omega}\sinh(\omega t)\\
-\frac{k_y}{\omega}\sinh(\omega t) &
\cosh(\omega t)-\frac{k_x}{2\omega}\sinh(\omega t)
\end{array}\right],
\eeq
$\omega^2=k_x^2/4-k_xk_y/\gamma$ if $k_x>4k_y/\gamma$, and the same expression with $\sinh\rightarrow \sin$, $\cosh\rightarrow \cos$, $\omega^2=k_xk_y/\gamma-k_x^2/4$, otherwise.
The fluctuations of $x$ are given by:
\beq
\begin{split}
\delta x(t)&=\frac{k_x}{u_0}\int_{-\infty}^tdt'\, e^{-\frac{k_x}{2}(t-t')}\left[\left(\cosh(\omega t)-\frac{k_x}{2\omega}\sinh(\omega t)\right)\left(\cosh(\omega t')+\frac{k_x}{2\omega}\sinh(\omega t')\right)
+\frac{k_xk_y}{\gamma\omega^2}\sinh(\omega t)\sinh(\omega t')\right]\delta u(t'),\nonumber\\
&=\frac{k_x}{u_0}\int_{-\infty}^tdt'\, e^{-\frac{k_x}{2}(t-t')}\left[\cosh\omega(t-t')-\frac{k_x}{2\omega}\sinh \omega(t-t')\right]\delta u(t'),
\end{split}
\eeq
and the average uncertainty:
\beq
\<(\delta x)^2\>=\frac{gk_x^2}{k_+c_0(1+\<\tau_b\>k_+c_0)}\int_{-\infty}^tdt'\, e^{-k_x(t-t')}{\left[\cosh\omega(t-t')-\frac{k_x}{2\omega}\sinh \omega(t-t')\right]}^2= \frac{gk_x}{2 k_+c_0 (1+\<\tau_b\>k_+c_0)},
\eeq
\end{widetext}
where the last equality is valid at steady state. Exactly the same result is obtained when $k_x<4k_y/\gamma$.
Using Eqs.~\ref{u1} and \ref{readout1}, we derive the uncertainty of the ramp rate readout:
\beq\label{unnet2}
\frac{\<(\delta c_1)^2\>}{(c_0k_y/\gamma)^2}=\frac{gk_x}{2u_0}.
\eeq
In addition we can also evaluate the variance of $y$:
\beq
\delta y(t)=\frac{k_x}{u_0}\int_{-\infty}^tdt'\, e^{-k_x(t-t')/2}\frac{k_y}{\omega}\sinh \omega(t-t')\delta u(t'),
\eeq
and therefore:
\begin{widetext}
\beq
\<(\delta y)^2\>=\frac{gk_x^2}{k_+c_0(1+\<\tau_b\>k_+c_0)}{\left(\frac{k_y}{\omega}\right)}^2\int_{-\infty}^tdt'\, e^{-k(t-t')}{\left[\sinh \omega(t-t')\right]}^2=\frac{gk_y\gamma}{2 k_+c_0(1+\<\tau_b\>k_+c_0)},
\eeq
\end{widetext}
from which we obtain the uncertainty of the concentration readout, using Eqs.~\ref{u0} and \ref{readout3}:
\beq
\frac{\<(\delta c_0)^2\>}{(c_0)^2}=\frac{gk_y/\gamma}{2u_0}.
\eeq

\section{Distribution of intervals of signaling activity for a receptor at equilibrium}\label{appC}
We now prove that for any receptor in thermal equilibrium:
\beq
\frac{\<(\delta \tau)^2\>}{\<\tau\>^2}\geq 1,
\eeq
where $\tau$ is the duration of an interval of uninterrupted signaling activity. With the results from the previous section, this proves $g\geq 2$ for any continuously signaling receptor, unless free energy is consumed in the binding/unbinding cycle.
Previously and in the main text, the receptor was assumed to signal whenever the receptor was bound, and the receptor was inactive otherwise ($\tau$ is then equal to the bound duration $\tau_b$), but our proof allows for a more general definition of signaling activity.
We model the receptor by a Markovian system with $N$ distinct states. $M$ of these states are called active and form the subset $A\subset \{1,\ldots,N\}$. The remaining states belonging to the complementary subset $I=\{1,\ldots,N\}\backslash A$ are called inactive.

Our goal is to calculate the distribution of intervals of uninterrupted activity when the system is in equilibrium, {\em i.e.} intervals where the system remains in $A$. In particular, we will show that this distribution is a weighted sum of exponentials with positive weights:
\beq\label{possum}
P(\tau)=\sum_{k=1}^M w_k\lambda_k e^{-\lambda_k \tau},\quad w_k\geq 0\quad\textrm{ and }\quad\sum_{k=1}^Mw_k=1.
\eeq
This result generalizes the one proven in the supporting information of \cite{Tu:2008p7110}. In that proof, active states were paired one-to-one to inactive states (whence $N=2M$), and, among the possible transitions from active to inactive states, only the ones within these pairs were allowed. Our proof places no restrictions on the number of states or on transitions between states.

A consequence of Eq.~\ref{possum} is that the relative variance of intervals is bounded below by one. This lower bound is only attained for a single-step process ($M=1$) for which the distribution of intervals is a pure exponential. More precisely,
\beq
\frac{\<(\delta \tau)^2\>}{\<\tau\>^2}=\frac{\<\tau^2\>}{\<\tau\>^2}-1=\frac{2\sum_{k=1}^M w_k\frac{1}{\lambda_k^2}}{\left(\sum_{k=1}^M w_k\frac{1}{\lambda_k}\right)^2}-1\geq 1,
\eeq
where the last step follows from Jensen's inequality, which becomes an equality if and only if $M=1$.

We now prove Eq.~\ref{possum}.
Let $Q=\{q_{ij}\}$ denote the matrix of rates from state $j$ to state $i$, with $q_{ii}=-\sum_{j\neq i}q_{ji}$.
Call $\ket{p}=(p_1,\ldots,p_N)^{\rm T}$ the vector of equilibrium probabilities, which satisfies $Q\ket{p}=\ket{p}$.
Because of detailed balance, we have $q_{ij}p_j=q_{ji}p_i$.
With proper ordering of indices, $Q$ can be decomposed according to the active and inactive states:
\beq
Q=\left[\begin{array}{cc}Q_{AA}&Q_{AI}\\Q_{IA}&Q_{II}\end{array}\right].
\eeq
Note that in \cite{Tu:2008p7110}, $Q_{AI}$ and $Q_{IA}$ are square, diagonal matrices.
Assuming that the system enters an active state at time $t=0$, this state will be $i\in A$ with probability
\beq
\frac{[Q_{AI}\ket{p_I}]_i}{\bra{1_A}Q_{AI}\ket{p_I}},
\eeq
where $Q_{AI}$ is the submatrix of rates from inactive to active states,
$\ket{p_I}$ is the projection of $\ket{p}$ onto subset $I$, and $1_{A}$ is a vector of ones of dimension $M$. Starting from an active state $j$ at time $0$, the probability of still being active in state $i\in A$ at time $t$ (if $i\in A$), or alternatively to have exited the active states via the inactive state $i\in I$ at time $t'\leq t$ (if $i\in I$), is given by $X_{ij}$, where $X$ is the solution to the following Master equation:
\beq
\frac{dX}{dt}=Q^{\rm abs}X,\qquad X=
e^{tQ^{\rm abs}},
\eeq
where $Q^{\rm abs}$ is the same as $Q$ but with absorbing inactive states: $Q^{\rm abs}_{AI}=0$ and $Q^{\rm abs}_{II}=0$. Thus:
\bea
&Q^{\rm abs}&=\left[\begin{array}{cc}Q_{AA}&0\\Q_{IA}&0\end{array}\right]\\
\textrm{and}&
\quad e^{tQ^{\rm abs}}&=\left[\begin{array}{cc}e^{tQ_{AA}}&0\\Q_{IA}Q_{AA}^{-1}(e^{tQ_{AA}}-1)&1\end{array}\right].
\eea
Therefore the total probability of having exited the active states at time $t'\leq t$, after starting in $j\in A$ at $t=0$, is:
\beq
\left[\bra{1_I}Q_{IA}Q_{AA}^{-1}(e^{tQ_{AA}}-1)\right]_j,
\eeq
where $\bra{1_I}$ is a vector of ones of dimension $N-M$. Finally, the total probability of having exited the active state at time $t'\leq t$ after entering it at time $t=0$ is:
\beq
C(t)=\frac{\bra{1_I}Q_{IA}Q_{AA}^{-1}(e^{tQ_{AA}}-1) Q_{AI}\ket{p_I}}{\bra{1_A}Q_{AI}\ket{p_I}}.
\eeq
which yields the probability distribution of intervals of activity:
\beq
P(\tau)=\frac{d C}{dt}(\tau)=\frac{\bra{1_I}Q_{IA}e^{\tau Q_{AA}}Q_{AI}\ket{p_I}}{\bra{1_A}Q_{AI}\ket{p_I}}.
\eeq
To exploit the property of detailed balance, we symmetrize $Q_{AA}$ by defining $\tilde Q=\{q_{ij}\sqrt{p_j/p_i}\}_{i\in A,j\in A}$. $\tilde Q$ being symmetric, it can diagonalized in an orthonormal base:
\beq
\tilde Q=-\sum_{k=1}^{M}\lambda_k \ket{u^k}\bra{u^k},\quad \textrm{with }\lambda_k>0\quad\textrm{and}\quad\braket{u^k}{u^{k'}}=\delta_{kk'}.
\eeq
The diagonal form of $Q_{AA}$ thus reads:
\beq
Q_{AA}=-\sum_{k=1}^{M}\lambda_k \ket{v^{{\rm R},k}}\bra{v^{{\rm L},k}},
\eeq
where $v^{{\rm R},k}_i=u^k_i\sqrt{p_i}$ and $v^{{\rm L},k}_i=u^k_i/\sqrt{p_i}$ are the right and left eigenvectors of $Q_{AA}$.
$P(\tau)$ can now be rewritten as:
\beq
P(\tau)=\frac{\sum_{k}e^{-\lambda_k \tau}\bra{1_I}Q_{IA}\ket{v^{{\rm R},k}}\bra{v^{{\rm L},k}}Q_{AI}\ket{p_I}}{\bra{1_A}Q_{AI}\ket{p_I}}.
\eeq
We have:
\beq
\begin{split}
\bra{v^{{\rm L},k}}Q_{AI}\ket{p_I}&=\sum_{i\in I,j\in A}u^{k}_jp_j^{-1/2}q_{ji}p_i\\
&=\sum_{i\in I,j\in A}u^{k}_jp_j^{-1/2}q_{ij}p_j\\
&=\sum_{i\in I,j\in A}u^{k}_j\sqrt{p_j}q_{ij}\\
&=\bra{1_I}Q_{IA}\ket{v^{{\rm R},k}}\\
&:= A_k.
\end{split}
\eeq
and therefore:
\beq
P(t)=\frac{\sum_{k=1}^M A_k^2 e^{-\lambda_k t}}{\sum_{k=1}^M \frac{A_k^2}{\lambda_k}},
\eeq
which establishes Eq.~\ref{possum} with $w_k=(A_k^2/\lambda_k)/(\sum_{k'}A_{k'}^2/\lambda_{k'})$.





\bibliographystyle{apsrev}


\bibliography{chemonourl,other}


\end{document}